\documentclass[twocolumn,pra,aps,showpacs,superscriptaddress]{revtex4-1}
\usepackage[utf8]{inputenc}
\usepackage{url,psfrag,graphicx}
\usepackage{dcolumn}
\usepackage{amsmath,amssymb}
\usepackage{pstricks}
\usepackage{epsfig} % ,placeins}
\usepackage{hyperref}

\def\<{\left<}
\def\>{\right>}
\def\ket|#1>{\left|#1\right>}
\def\bra<#1|{\left<#1\right|}
\def\elem<#1|#2|#3>{\left<#1\right|#2\left|#3\right>}
\def\({\left(}
\def\){\right)}

\def\pmat#1{\begin{pmatrix}#1\end{pmatrix}}
\def\{{\left\lbrace}
\def\}{\right\rbrace}

\begin{document}

\title[Short Title]{Local Quantum Thermometry using Unruh-De Witt Detectors}

\author{Sandra Robles}
\affiliation{Instituto de Física Teórica, UAM/CSIC, Madrid, Spain.}
\affiliation{Departamento de Física Teórica, UAM, Madrid, Spain.}

\author{Javier Rodríguez-Laguna}
\affiliation{Departamento de Física Fundamental, UNED, Madrid, Spain.}
\affiliation{Instituto de Física Teórica, UAM/CSIC, Madrid, Spain.}

\date{February 7, 2017}

\begin{abstract}
We propose an operational definition for the local temperature of a
quantum field employing Unruh-DeWitt detectors, as used in the study
of the Unruh and Hawking effects. With this definition, an
inhomogeneous quantum system in equilibrium can have different local
temperatures, in analogy with the Tolman-Ehrenfest theorem from
general relativity. We study the local temperature distribution on the
ground state of hopping fermionic systems on a curved background. The
observed temperature tends to zero as the thermometer-system coupling
$g$ vanishes. Yet, for small but finite values of $g$, we show that
the product of the observed local temperature and the logarithm of the
local speed of light is approximately constant. Our predictions should
be testable on ultracold atomic systems.
\end{abstract}

\maketitle

%%%%%%%%%%%%%%%%%%%%%%%%%%%%%%%%%%%%%%%%%%%%%%%%%%%%%%%%%%%%%%%%%%%

\section{Introduction}

Recently, quantum simulators built upon ultracold atomic gases
\cite{Lewenstein.book} have been designed in order to explore the very
interesting interplay between quantum mechanics and curved space-time
\cite{Boada.11}, including the effects of dimensionality
\cite{Boada.12} or unusual topology \cite{Boada.15}. Moreover, a
detailed proposal for a quantum simulator to explore Unruh physics in
cold atoms has been put forward \cite{Laguna.16}. The idea behind all
the proposed quantum simulators on curved space-times
\cite{Lewenstein.book, Boada.11, Boada.12, Boada.15, Laguna.16} is the
following: a static metric with an inhomogeneous time-lapse function
$|g_{00}(x)|^{1/2}$ for fermionic systems can be simulated by tuning
the local hopping amplitudes between the cells of an optical
lattice. This relation can be also understood in reverse: an
inhomogeneity in the hopping amplitudes may be read as a non-trivial
space-time metric. This idea has sparked interest in the low energy
states of these inhomogeneous spin chains and fermionic hopping
models, which can be understood as dynamics on a curved metric. For
example, it has been shown how a modulation of the metric can give
rise to ground states (GS) which present extremely long-range
correlations, such as the {\em rainbow state}
\cite{Vitagliano.10,Ramirez.14}. The entanglement entropy of the GS of
local quantum systems usually follows the {\em area law}
\cite{Sredniki.93,Eisert.10}, but in a curved metric we can have a
strong violation, with a volumetric growth of the block
entropies. This led to a thermal interpretation of the rainbow state
\cite{Ramirez.15}, which can be viewed as a thermo-field double. Thus
we see that, in some situations, it makes sense to attach a non-zero
temperature to a quantum ground state.

One of the most surprising results in thermodynamics on curved
space-times was stated by Richard Tolman and Paul Ehrenfest in 1930
\cite{Tolman.Ehrenfest,Takagi.85}: the temperature of an equilibrium
system in a static space-time may vary from point to point, and it is
inversely proportional to the local lapse function,

\begin{equation}
T(x) \cdot |g_{00}(x)|^{1/2} = \text{const.}
\label{eq:tolman_ehrenfest}
\end{equation}
The result is of thermodynamical nature, and can be proved without any
assumptions on the dynamics \cite{Rovelli.10}. It can be applied to
the study of the {\em Unruh effect}: an accelerated observer traveling
through a Minkowski vacuum must feel a thermal bath of particles at a
temperature proportional to its acceleration
\cite{Unruh.76,Birrell.Davies}. Due to the principle of equivalence,
such an observer can be considered to be at rest in {\em Rindler
  space-time}, which is characterized by a lapse function which
increases linearly with the distance to a {\em horizon},
$|g_{00}(x)|^{1/2}\propto x$. Then, the Tolman-Ehrenfest theorem
predicts that the local temperature must decay as the inverse of that
same distance, $T(x)\propto x^{-1}$ \cite{Footnote.1,Verch}. It is relevant
to notice that the Unruh effect is defined in an operational way
\cite{Takagi.85,Birrell.Davies}: an {\em Unruh-DeWitt} detector is
defined as a simple quantum system with a local monopolar interaction
with the field. The temperature will manifest itself in the quantum
fluctuations within the detector \cite{Footnote.2}.

A great amount of theoretical work has been devoted to the {\em
  locality-of-temperature} problem, i.e., to find under which
conditions a subsystem of a global system at temperature $T$ can be
considered to be again in a thermal state at the same temperature
\cite{Eisert.14,Acin.15,Fazio.15}. In general terms, the answer is
that this is possible when a certain measure of the energy contained
in the correlations is lower than the physical temperature, $T$. Thus,
this work will explore the opposite limit, when $T=0$, so quantum
correlations can create non-trivial local thermal effects.  Thus, one
may ask how small a thermometer can be in order to make sensible
measurements. Quantum thermometry is indeed an area undergoing a rapid
growth. The idea of using a single qubit as a thermometer has been put
forward recently by several groups
\cite{Jevtic.15,Sanpera_PRL.15,Oliveira.15,Paris.11}. In this case,
the fluctuations in the temperature estimate should be taken into
account \cite{Uffink.99,Sanpera_NJP.15,Paris.15}, which are expected
to follow the Landau relation, $\Delta T\sim T^2/C$, where $C$ is the
heat capacity of the system.

This work proposes to explore local quantum thermometry on the ground
state of inhomogeneous free fermionic Hamiltonians by observing the
quantum fluctuations of a single-qubit Unruh-DeWitt detector, locally
linked to our system. The long term average of the occupation provides
an estimate of the local temperature, while its frequency dependence
provides further information about the system. We show that, for
finite couplings between the detector and the system, the observed
local temperature and the time-lapse are related via a modification of
the Tolman-Ehrenfest relation. Nonetheless, when the coupling tends to
zero, the temperature vanishes, at it should on a ground state. The
reason to employ free systems is that we will focus on the interplay
between geometry and thermal effects, and we leave the effects of
interaction for further work. Notice that, despite of our use of the
Unruh-De Witt detector, our measurement does not bear relation to the
Unruh effect.

There have been other proposals to define effective and local
temperatures for non-equilibrium and/or inhomogeneous systems in the
literature. Some of the most relevant are based either on the
fluctuation-dissipation theorem
\cite{Cugliandolo.97,Cugliandolo.11,Foini.11} or the connection to a
thermal bath with a vanishing heat flow
\cite{Engquist.81,Dubi.09,Caso.10,Caso.11,Eich.16,Shastry.15}. We will
comment on the relation to our approach at the end of this work.

This article is organized as follows. In section \ref{sec:model} we
describe our physical model, an Unruh-DeWitt detector locally attached
to the ground state of a fermionic system on a curved background. The
methodological issues are discussed in section \ref{sec:calc}, and the
numerical results are shown in section \ref{sec:numerical}. Section
\ref{sec:sqd} is devoted to a variational general study of the physics
of single-qubit detectors in interaction with free fermionic
systems. This article ends in section \ref{sec:conclusions} with a
summary of the conclusions and suggestions for further work.

%%%%%%%%%%%%%%%%%%%%%%%%%%%%%%%%%%%%%%%%%%%%%%%%%%%%%%%%%%%%%%%%%%%

\section{Unruh-DeWitt Thermometry}
\label{sec:model}

Let us consider a system of spinless fermions on $L$ sites
characterized by a Hamiltonian $H_S$, and let $c^\dagger_i$ denote the
creation operator at site $i$. We introduce a new site, the
Unruh-DeWitt detector or thermometer, with label 0 and a chemical
potential $\mu>0$, whose Hamiltonian is:

\begin{equation}
H_D = \mu c^\dagger_0 c_0.
\label{eq:dewitt}
\end{equation}

\begin{figure}
\includegraphics[width=8cm]{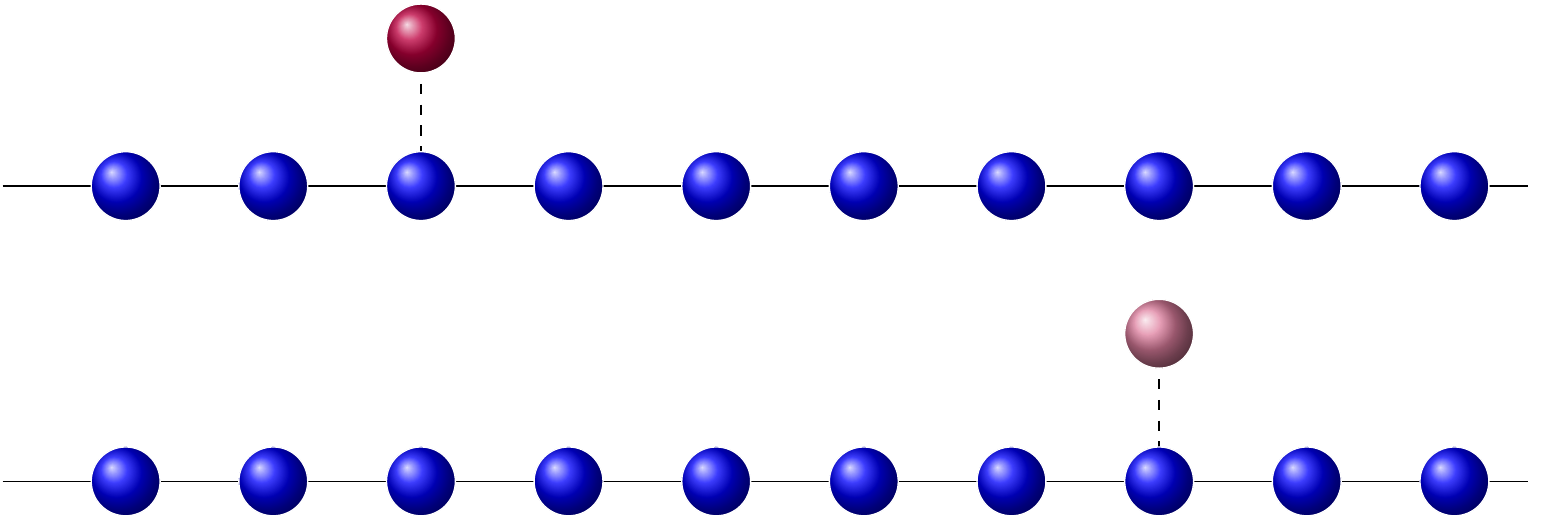}
\caption{We set up a fermionic chain with $L$ sites in its ground
  state (blue), plus a {\em thermometer site} or Unruh-DeWitt detector
  (red), initially empty and uncoupled. At time $t=0$ we establish a
  local hopping between them (dashed line), and trace the evolution of
  the expected occupation of the thermometer as a function of time,
  $n_0(t)$. Since the occupied and empty states have different
  energies, we can {\em infer} a temperature from the long term
  behavior of $n_0(t)$. The inferred temperature may depend on the
  position of the thermometer, as illustrated in the panels.}
\label{fig:illust}
\end{figure}

Let $H_0\equiv H_S+H_D$, and let us cool the system into its ground
state, which will contain $L/2$ fermions (half-filling) in the system
while the detector will be empty. We now quench the system by
attaching the detector to site $p$ of the system via an interaction
term of the form

\begin{equation}
H_I = g \(c^\dagger_0 c_p + h.c.\),
\label{eq:interaction}
\end{equation}
where $g$ is a (small) coupling constant. The total Hamiltonian of the
system is now given by

\begin{equation}
H=H_0 + H_I = H_S+H_D+H_I,
\label{eq:total}
\end{equation}
see Fig. \ref{fig:illust} for an illustration. If the detector is in
the extreme, the system presents some similarity to a Kondo lattice,
but it remains a pure hopping Hamiltonian, non-interacting.

After the quench, we observe that the expected value of the occupation
of the detector is a function of time, $\<n_0(t)\>$, and we can define
$n_0$ to be its long-term time average,

\begin{equation}
n_0\equiv \lim_{\tau\to\infty} {1\over \tau} \int_0^\tau dt\,\<n_0(t)\>,
\label{eq:def_n0}
\end{equation}
if this limit exits. Since the energy difference between the empty and
the occupied states of the detector is $\mu$, which we assume to be
sufficiently above the Fermi energy of the system, we can give a
thermal interpretation to that magnitude:

\begin{equation}
n_0 \equiv {1\over 1+\exp(\beta\mu)},
\label{eq:fermi_dirac}
\end{equation}
from which we can infer a local temperature $T=1/\beta$, associated to
site $p$. If the energy provided by the coupling, $\sim g$, is small,
we can assume that we are not perturbing the system noticeably and,
therefore, we are measuring an intrinsic property of the quantum
system. Of course, the proper value of the temperature should always
be taken as $g\to 0$. For finite values of $g$, we will speak of {\em
  observed} values of the local temperature.

Notice that this procedure bears a strong similarity to the
operational definition of the Unruh temperature \cite{Birrell.Davies},
the main difference being that our detector is at rest.

%%%%%%%%%%%%%%%%%%%%%%%%%%%%%%%%%%%%%%%%%%%%%%%%%%%%%%%%%%%%%%%%%%%%%%%

\section{Computing the thermometer occupation}
\label{sec:calc}

For concreteness, let us consider our system to be a 1D free fermion
lattice with $L$ sites and a position-dependent hopping amplitude:

\begin{equation}
H_S=-\sum_i t_i c^\dagger_i c_{i+1} + h.c. \ ,
\label{eq:basic_model}
\end{equation}
where the $t_i$ are the hopping amplitudes, encoding the geometry. If
they are slowly varying, they can be understood as a local time-lapse
function $|g_{00}(x)|^{1/2}$ of a static metric \cite{Boada.11}:

\begin{equation}
ds^2 = -t^2(x) dt^2 + dx^2,
\label{eq:metrica}
\end{equation}
where we assume $x_i=i \Delta x$ and $t(x_i)\approx t_i/\Delta x$. We
can also think of $t(x)$ as a local speed of light in an optical
metric. Notice that the restriction to a 1D non-interacting system is
only made for convenience. An important property of the Hamiltonian
\eqref{eq:basic_model} is that its single-particle spectrum presents
particle-hole symmetry. Thus, for the ground state at half-filling,
the particle density $\<n_i\>=1/2$ is always homogeneous.
\vspace{1mm}

Let us compute the local temperature defined by
Eq. \eqref{eq:fermi_dirac}. Before the quench the Hamiltonian is
$H_0=H_S+H_D$, and after the quench it is $H=H_0+H_I$. Both
Hamiltonians are free, thus their eigenstates can be obtained in terms
of single-body energies and orbitals:

\begin{align}
H_0 \to\qquad& \{ \epsilon_k,\qquad b^\dagger_k=\sum_i B_{ki}
c^\dagger_i\}, \\
H \to\qquad &\{\eta_l,\qquad d^\dagger_l=\sum_i D_{li}
c^\dagger_i\}.
\label{eq:orbitals}
\end{align}

The linear transformations among the single-body orbitals
$b^\dagger_k$, $d^\dagger_k$ and $c^\dagger_i$ are all
unitary. Furthermore, we define

\begin{equation}
d^\dagger_l \equiv \sum_k U_{lk} b^\dagger_k = \sum_{k,i} D_{li} \bar
B_{ik} b^\dagger_k.
\label{eq:d.from.b}
\end{equation}

The initial state is the ground state of $H_0$:

\begin{equation}
\ket|\Psi_0>=\prod_{k\in K} b^\dagger_k \ket|0>,
\label{eq:initial.state}
\end{equation}
where $K$ is the set of occupied levels in the initial system, i.e.:
those whose energy $\epsilon_k<0$ (we will assume it to be
non-degenerate, so there are no zero modes). Let us express the time
evolution in the Heisenberg image, making the operators evolve. Thus,
we need to obtain

\begin{equation}
n_0(t)\equiv \bra<\Psi_0| c^\dagger_0(t) c_0(t) \ket|\Psi_0>.
\label{eq:exp.n0}
\end{equation}

The orbitals of $H$ evolve as $d^\dagger_k(t)=d^\dagger_k e^{-i \eta_k
  t}$, where $d^\dagger_k(0)=d^\dagger_k$. The evolution of the
on-site $c^\dagger_i(t)$ operators is given by:

\begin{equation}
c_0^\dagger(t)=\sum_l \bar D_{0l} d^\dagger_l e^{-i \eta_k t}.
\label{eq:ev.c0}
\end{equation}
Putting all together we obtain

\begin{align}
n_0(t)&=\sum_{l,l'} \bar D_{0l} D_{0l'} e^{i \eta_l t} e^{-i \eta_{l'}
  t} \bra<\Psi_0|d^\dagger_l d_{l'} \ket|\Psi_0>\\
&=\sum_{l,l'} \bar D_{0l} D_{0l'} e^{-i (\eta_l-\eta_{l'}) t} 
\sum_{k\in K} U_{lk} \bar U_{l'k}.
\label{eq:evol.n0}
\end{align}

From here we read that the Fourier transform of the temporal
fluctuations of the detector occupation, $\hat n_0(\omega)$, has peaks
at frequencies $\omega_{ll'}\equiv \eta_l-\eta_{l'}$:

\begin{equation}
\hat n_0(\omega)= \sum_{l,l'} W_{ll'}\, \delta(\omega-\omega_{ll'}),
\label{eq:n0_fourier}
\end{equation}
with weights given by the expression:

\begin{equation}
W_{ll'}=\sum_{k\in K} \bar D_{0l} D_{0l'} U_{lk} \bar U_{l'k}.
\label{eq:weights}
\end{equation}

Assuming that the $\eta_l$ are all different, we can read the
expression for the long-term average of the expectation value of the
occupation, \eqref{eq:def_n0}, as the zero-frequency component:

\begin{equation}
n_0 = \sum_l W_{ll} = \sum_l \sum_{k\in K} |D_{0l}|^2 |U_{lk}|^2.
\label{eq:longtermaverage}
\end{equation}

For a finite system, expression \eqref{eq:longtermaverage} always
makes sense and converges to the long term average of the occupation
as long as there are no degeneracies in the single-particle spectrum
of $H$, $\{\eta_l\}$. A relevant question in practice is what does
{\em long term} mean exactly. The answer is: long enough for all
non-zero frequencies in expression \eqref{eq:n0_fourier} to average
out, which will require a time inversely proportional to the slowest
non-zero value of $\eta_l-\eta_{l'}$.

%%%%%%%%%%%%%%%%%%%%%%%%%%%%%%%%%%%%%%%%%%%%%%%%%%%%%%%%%%%%%%%%%%%%%

\section{Numerical Results}
\label{sec:numerical}

We have performed numerical simulations in order to explore the
relation between the local temperature, the thermometer occupation and
the local properties of the state. In all cases, unless otherwise
specified, we choose the thermometer chemical potential $\mu=0.5$ and
$g=0.1$. 

\begin{figure}
\includegraphics[width=8.5cm]{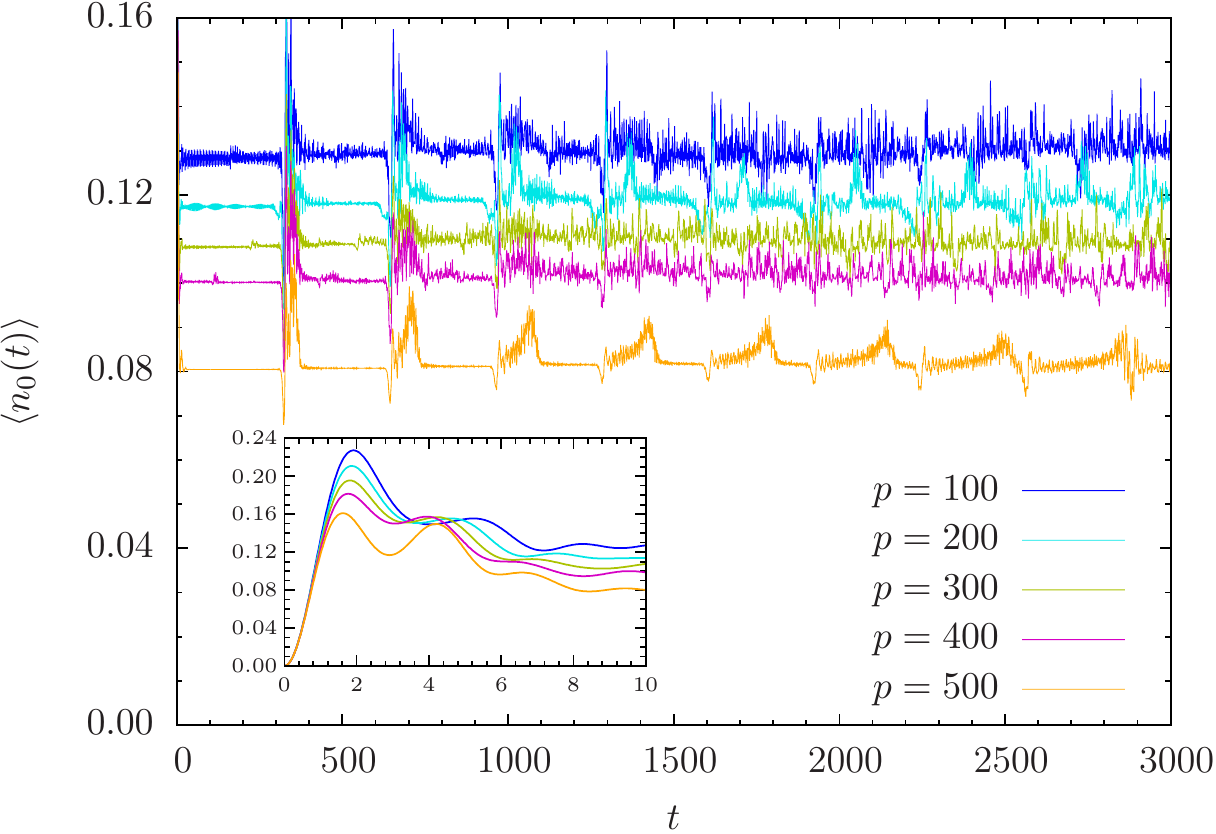}
\caption{Time evolution of the expected value of $\<n_0(t)\>$, i.e.,
  the occupation of the thermometer site, for a fermionic
  inhomogeneous hopping model with Rindler-like metric and $L=500$ and
  couplings $t_i=0.6+0.4 (i/L)$, using $\mu=0.55$ and $g=0.6$. The
  inset shows the same values for shorter times. Notice that the
  initial values, $\<n_0(0)\>=0$ in all cases, but it jumps to a high
  level in a very short time.}
\label{fig:rindler_quench}
\end{figure}

In Fig. \ref{fig:rindler_quench} we show the time evolution of the
expected value of the occupation of the thermometer $\<n_0(t)\>$ when
it is attached to different sites of a $L=500$ fermionic Rindler-like
chain with couplings of the form $t_i=t_0 + i\Delta t$ ($t_0=0.6$ and
$\Delta t=0.4$) and open boundaries. Notice that the different values
of the long-time average are easy to spot from the beginning, and
rather marked. The periodic bursts are related to the time taken by
the perturbation created by the quench to bounce back at the
boundaries and return.

\begin{figure}
\includegraphics[width=8cm]{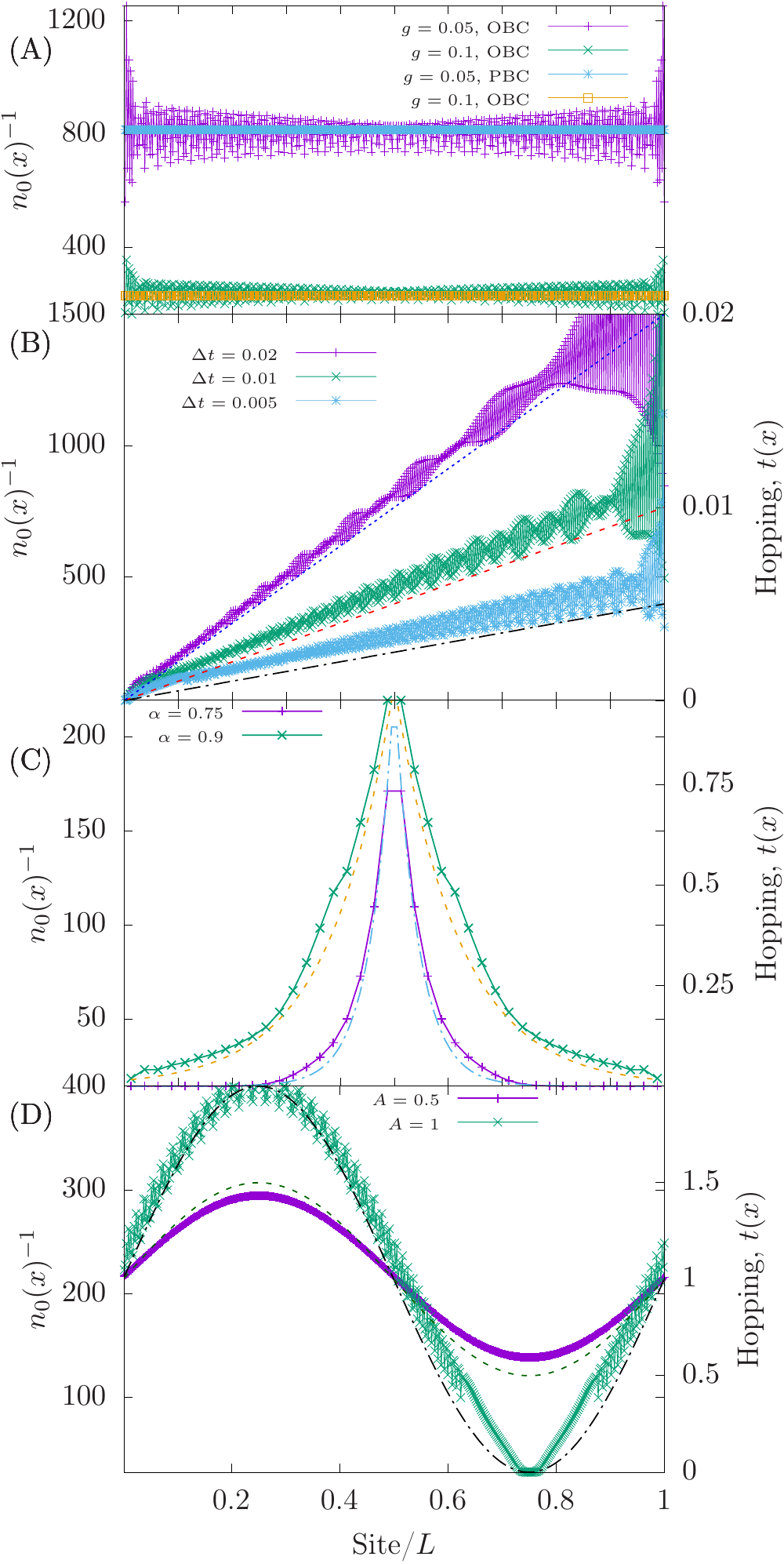}
\rput(-8.5,15.5){(A)}
\rput(-8.5,12.5){(B)}
\rput(-8.5,8.5){(C)}
\rput(-8.5,4.5){(D)}
\caption{Inverse of the average local thermometer occupation,
  $n_0(x)^{-1}$ for different background geometries. When suitable,
  the hopping distribution is shown in dotted lines. (A)
  Homogeneous system with open and periodic boundary conditions (OBC
  and PBC) and two different values of $g$, for a system with $L=500$
  sites and $\mu=0.5$. (B) Rindler geometry, Eq. \eqref{eq:rindler}
  with $L=500$, $\mu=0.5$ and $g=0.1$, using $\Delta t=0.005$, $0.01$
  and $0.02$. (C) Rainbow geometry, Eq. \eqref{eq:rainbow}, with
  $L=40$, $\mu=0.5$, $g=0.1$ and two values of $\alpha=0.9$ and
  $0.7$. (D) Sinusoidal geometry, Eq. \eqref{eq:sine}, with $L=500$,
  $\mu=0.5$, $g=0.1$ and $t_0=1$, with two different amplitudes: $A=1$
  and $A=1/2$.}
\label{fig:tempsite}
\end{figure}
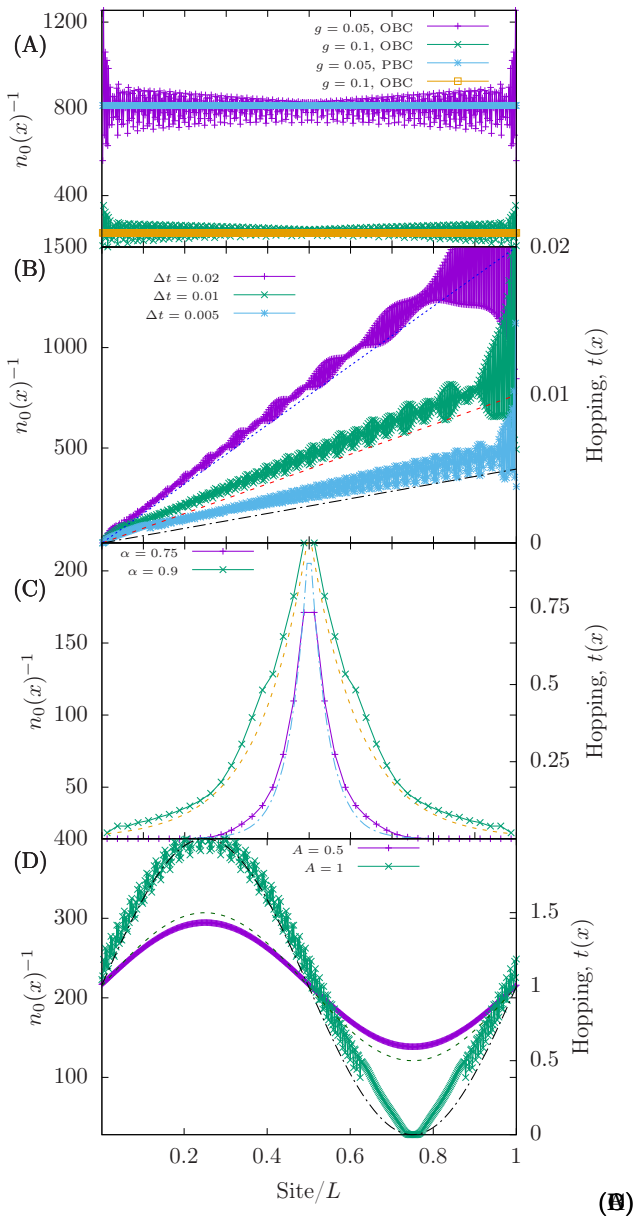

Fig. \ref{fig:tempsite} shows the inverse of the long term average of
the occupation of the thermometer when attached at different sites,
$n_0(x)^{-1}$, obtained using Eq. \eqref{eq:longtermaverage}, for
different background geometries, which we will describe from top to
bottom. (A) A constant hopping term, $t_i=1$, both with open and
periodic boundary conditions (OBC and PBC) for a system with
$L=500$. For PBC, the occupation is homogeneous due to the translation
invariance. For OBC, the average value of $n_0^{-1}$ is the same as
for PBC, but we observe large fluctuations due to the boundaries. (B)
Rindler chain,

\begin{equation}
t_i=i\Delta t,
\label{eq:rindler}
\end{equation}
with fixed $\Delta t$ and open boundaries. The left extreme of the
system, $t\sim 0$, behaves similarly to a horizon. We use also
$L=500$, $\Delta t=0.005$, $0.01$ and $0.02$, and $g=0.1$. In this
case, the result is more surprising: we observe that $n_0(x)^{-1} \sim
x$, in similarity to the growth of the hopping term. Thus, we can
assert our main conjecture:

\begin{equation}
n_0(x)^{-1} \sim t(x),
\label{eq:ourtolman}
\end{equation}
where the proportionality constant between them may depend on the
parameters of the thermometer, $g$ and $\mu$. This expression,
nonetheless, is only approximate. Moreover, the local occupation of
the thermometer presents strong parity oscillations. (C) Rainbow
chain,

\begin{equation}
t_i=\alpha^{|i-L/2|},
\label{eq:rainbow}
\end{equation}
with $\alpha\in (0,1]$, i.e., the hoppings fall exponentially from the
center. The ground state of this system presents volumetric growth
of the entanglement \cite{Vitagliano.10,Ramirez.14,Ramirez.15}, and
can be interpreted as a thermo-field state. In this case, using
$L=40$ and $\alpha=0.9$ and $0.7$, we also observe the conjectured
form \eqref{eq:ourtolman} to hold approximately. In this case, no
parity oscillations appear, but we can see that the occupation
saturates when we move away from the center. (D) Sinusoidal chain,

\begin{equation}
t_i=1+A\sin(2\pi i /L),
\label{eq:sine}
\end{equation}
which we explore for $L=500$ and $A=0.5$ and $1$. The first case
follows our conjectured form \eqref{eq:ourtolman} very accurately. The
second, $A=1$ presents a horizon at $i=3L/4$, and around its
neighborhood our conjecture is less accurate.

\begin{figure}
\includegraphics[width=8cm]{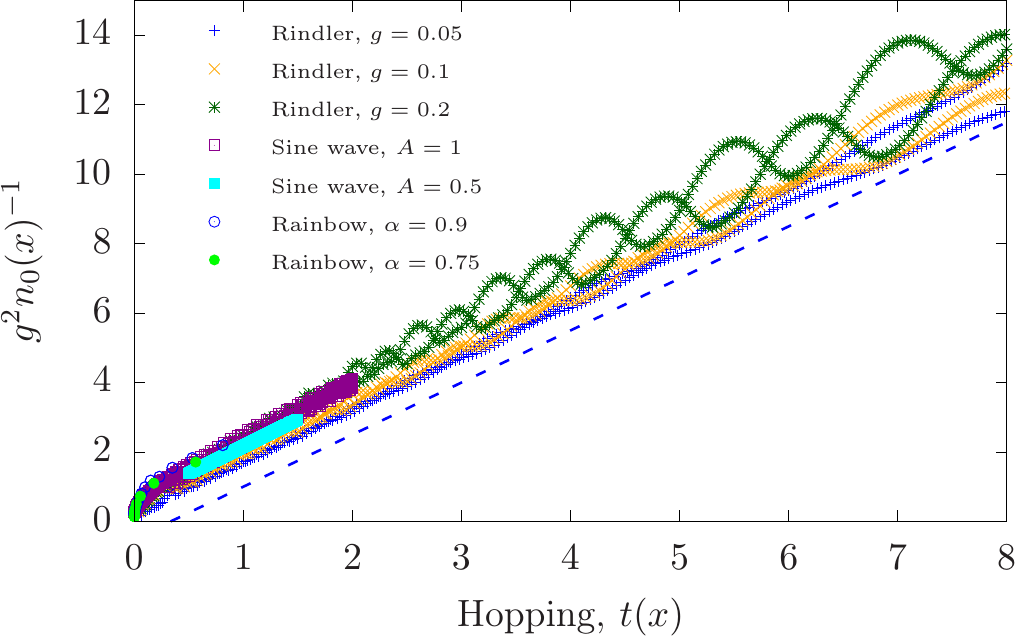}
\includegraphics[width=8cm]{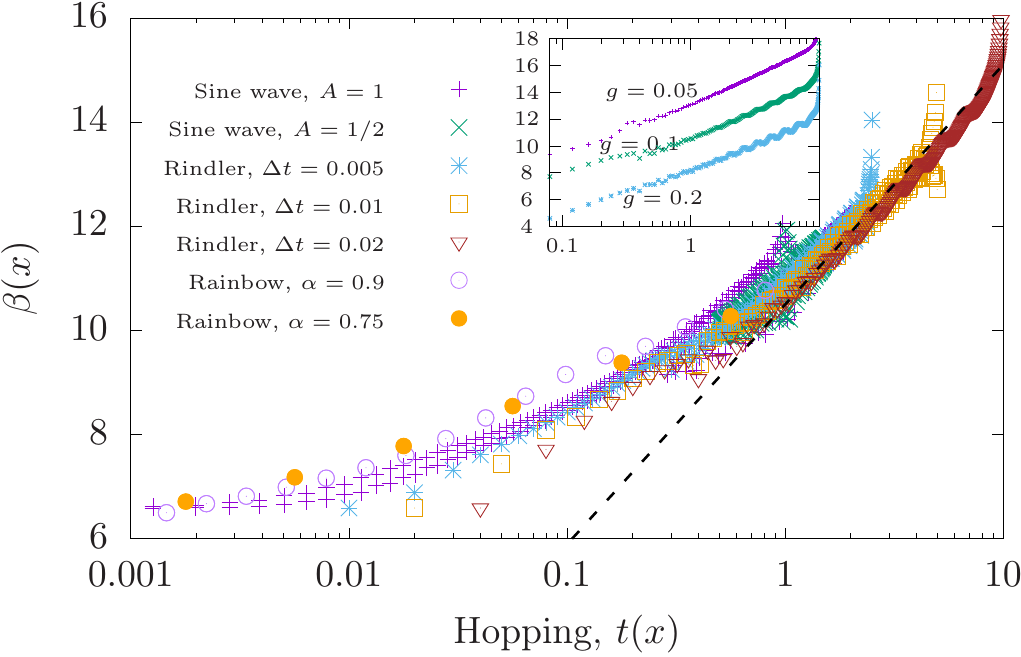}
\caption{Top: Plot of $g^2n_0(x)^{-1}$ versus the hopping $t(x)$, for
  different geometries and values of $g$, showing the approximate
  linear relationship. Concretely: Rindler system with $\Delta t=0.02$
  and $g=0.05$, $0.1$ and $0.2$, and the rainbow and sinusoidal
  systems shown in Fig. \ref{fig:tempsite}. The dotted straight line
  has slope $1.618$ and is slightly shifted for clarity. Bottom: Local
  inverse temperature, $\beta(x)$ plotted against the local hopping
  $t(x)$ for some of the systems shown in Fig. \ref{fig:tempsite},
  always using $g=0.1$. The hopping axis is shown in log-scale, in
  order to highlight the nearly logarithmic behavior of the relation
  between $\beta(x)$ and $t(x)$ for large $t$, see
  Eq. \eqref{eq:ourtolman3}. Notice the approximate data collapse to a
  single curve. The dotted line is $10+2\log(t)$. Inset: effect of
  $g$, shown plotting also $\beta$ vs $t$ in logarithmic scale, for a
  Rindler system with $\Delta t=0.02$ and 500 sites, for three values
  of $g$: 0.05, 0.1 and 0.2.}
\label{fig:temphop}
\end{figure}

The local relation between occupation $n_0$ and hopping $t$ is further
explored in the top panel of Fig. \ref{fig:temphop}, which plots
$g^2n_0$ vs $t$ for different Rindler systems, varying $g$, and the
sine and rainbow systems. The data seem to collapse to a straight
line, which amounts to an improvement of our previous relation
\eqref{eq:ourtolman} to

\begin{equation}
n_0(x)^{-1} \sim t(x)/g^2.
\label{eq:ourtolman2}
\end{equation}
The physical reason for the $g^2$ factor will be explained in the next
section. Furthermore, assuming Eq. \eqref{eq:ourtolman2} to be true,
we may also conjecture that the local inverse temperature
$\beta(x)\equiv T^{-1}(x)$ will behave like

\begin{equation}
\beta(x) \sim \log(t(x)/g^2),
\label{eq:ourtolman3}
\end{equation}
and this expression is tested in Fig. \ref{fig:temphop}, which shows
the local hopping in the horizontal axis, in logarithmic scale, and
$\beta(x)$ in the vertical one, for most of the systems used in
Fig. \ref{fig:tempsite}, using always $g=0.1$. For large $t$ the
relation between $\beta$ and $t$ is shown to be approximately
logarithmic, and for the whole range of values considered they seem to
collapse to a single curve. The effect of varying $g$ on the inverse
temperature is shown in the inset: it amounts to a vertical additive
shift, as it should be apparent from Eq. \eqref{eq:ourtolman3}.

\begin{figure}
\includegraphics[width=9cm]{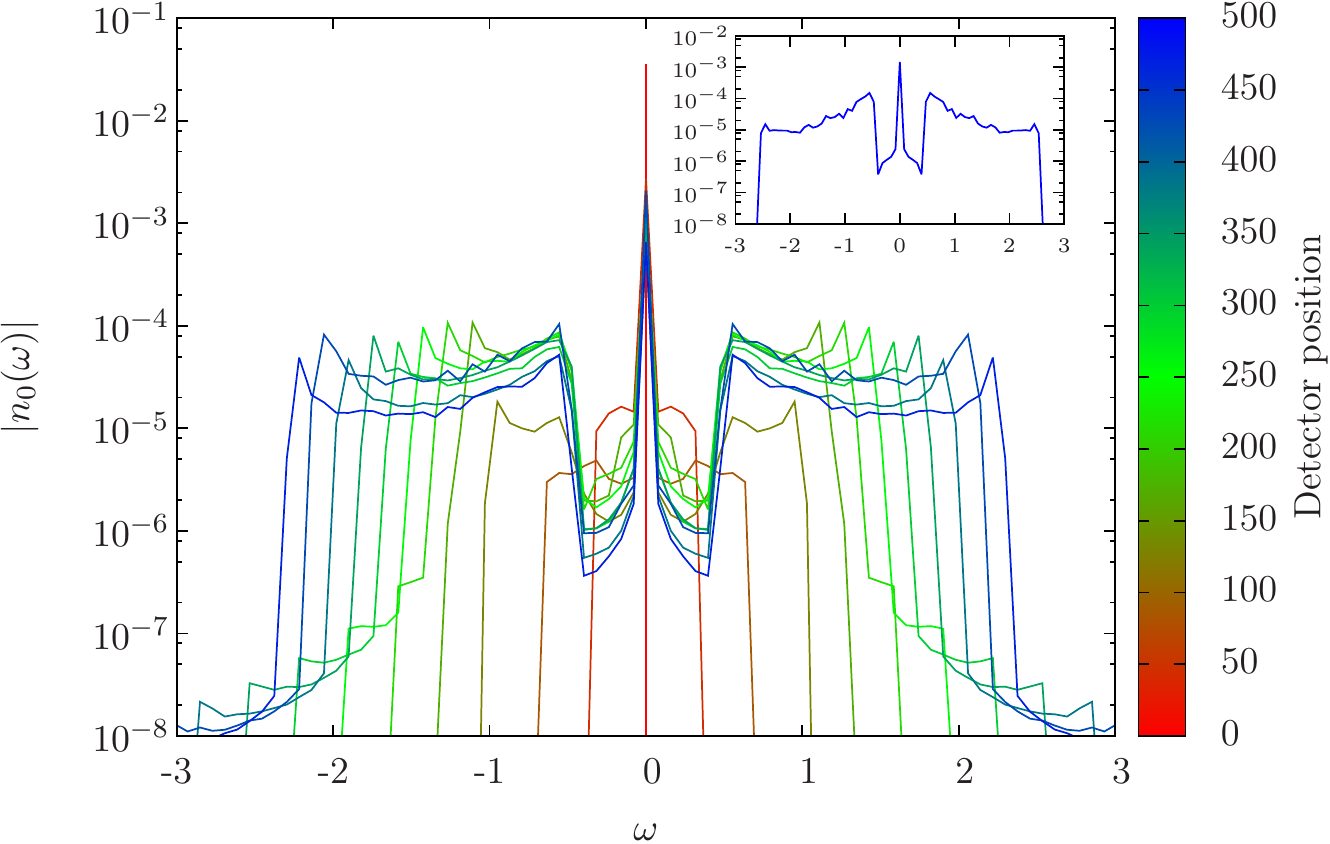}
\caption{Spectral decomposition of the thermometer noise,
  $|\<\hat n_0(\omega)\>|$ when the detector is placed at different
  positions over a Rindler system with $L=500$. Inset: spectral
  decomposition of the thermometer noise when the detector is placed
  on a homogeneous system.}
\label{fig:spectrum}
\end{figure}

The fluctuations in the thermometer occupation can be analysed beyond
their long-term average value. The full spectral decomposition of
$\<n_0(t)\>$ can be studied using Eq. \eqref{eq:n0_fourier}. In
Fig. \ref{fig:spectrum} we show the frequency decomposition of the
quantum noise on the thermometer, $|\<\hat n_0(\omega)\>|$ for a Rindler
system with $t_i=i/L$ and $L=500$, $g=0.1$ and $\mu=0.5$, when the
Unruh-DeWitt detector is placed at different sites. Notice that the
central peak, which corresponds to the long-term average $n_0$, is
relatively isolated. The active frequencies correspond to a block
which gets broader as we move away from the horizon.

For comparison, the inset of Fig. \ref{fig:spectrum} shows the same
spectral decomposition $|\<\hat n_0(\omega)\>|$ for the quantum noise of
the detector at any point of a homogeneous system. The shape is rather
similar to the response functions for Rindler space: the isolated
central peak plus the continuous block of frequencies.

%%%%%%%%%%%%%%%%%%%%%%%%%%%%%%%%%%%%%%%%%%%%%%%%%%%%%%%%%%%%%%%%%%%%%%

\section{Single Qubit Detectors}
\label{sec:sqd}

Let us discuss how the the single-body spectrum of a free fermionic
system changes when a new site is attached to site $p$, as shown in
Eq. \eqref{eq:interaction}, which we will call a {\em single qubit
  detector} (SQD), see Fig. \ref{fig:illust} for an illustration. Let
the unperturbed system be characterized by a set of single-body
orbitals $\{\psi^k_i\}$, with energies $E_k$.

A simple yet very accurate study can be done using a two-level
variational approach, in which each deformed single-body state is
obtained minimizing the energy within the subspace spanned by the
original orbital and the state localized in the new site. For each
unperturbed orbital, $k$, we propose an Ansatz of this form:

\begin{equation}
\ket|\Psi>_k= \alpha_k \ket|1>_k\otimes\ket|0>_D + \beta_k
\ket|0>_k\otimes\ket|1>_D,
\label{eq:ansatz_pimple}
\end{equation}
where $\{\ket|0>_k,\ket|1>_k\}$ denote the states in which mode $k$ is
either empty or occupied, and the same reads for
$\{\ket|0>_D,\ket|1>_D\}$ and the detector. The effective Hamiltonian
of this two-level system can be written as:

\begin{equation}
H_{eff}=\pmat{ E_k & g\psi^k_p \\ \bar g \bar\psi^k_p & \mu }.
\label{eq:effective_ham_pimple}
\end{equation}
Notice that only $\psi^k_p$ is relevant in this approach. The energy
shift for the orbital will be given by

\begin{align}
\tilde E_k &= {1\over 2}\( E_k+\mu \pm
\sqrt{(E_k-\mu)^2+4 g^2|\psi^k_p|^2} \) \\
&\approx  E_k + {g^2|\psi^k_p|^2\over E_k-\mu}.
\label{eq:deformed_energy}
\end{align}
Notice that the expression presents a pole at $E_k=\mu$, although we
will stay safe: $\mu$ is always chosen to be sufficiently above the
Fermi energy, which is zero in our case. Correspondingly, the
probability of finding the fermion in the new site is now

\begin{equation}
|\beta_k|^2 \approx { (\tilde E_k-E_k)^2 \over (\tilde E_k-E_k)^2 +
  g^2|\psi^k_p|^2 } \approx {g^2|\psi^k_p|^2\over (E_k-\mu)^2}.
\label{eq:deformed_occupation}
\end{equation}

\begin{figure}
\includegraphics[width=8.5cm]{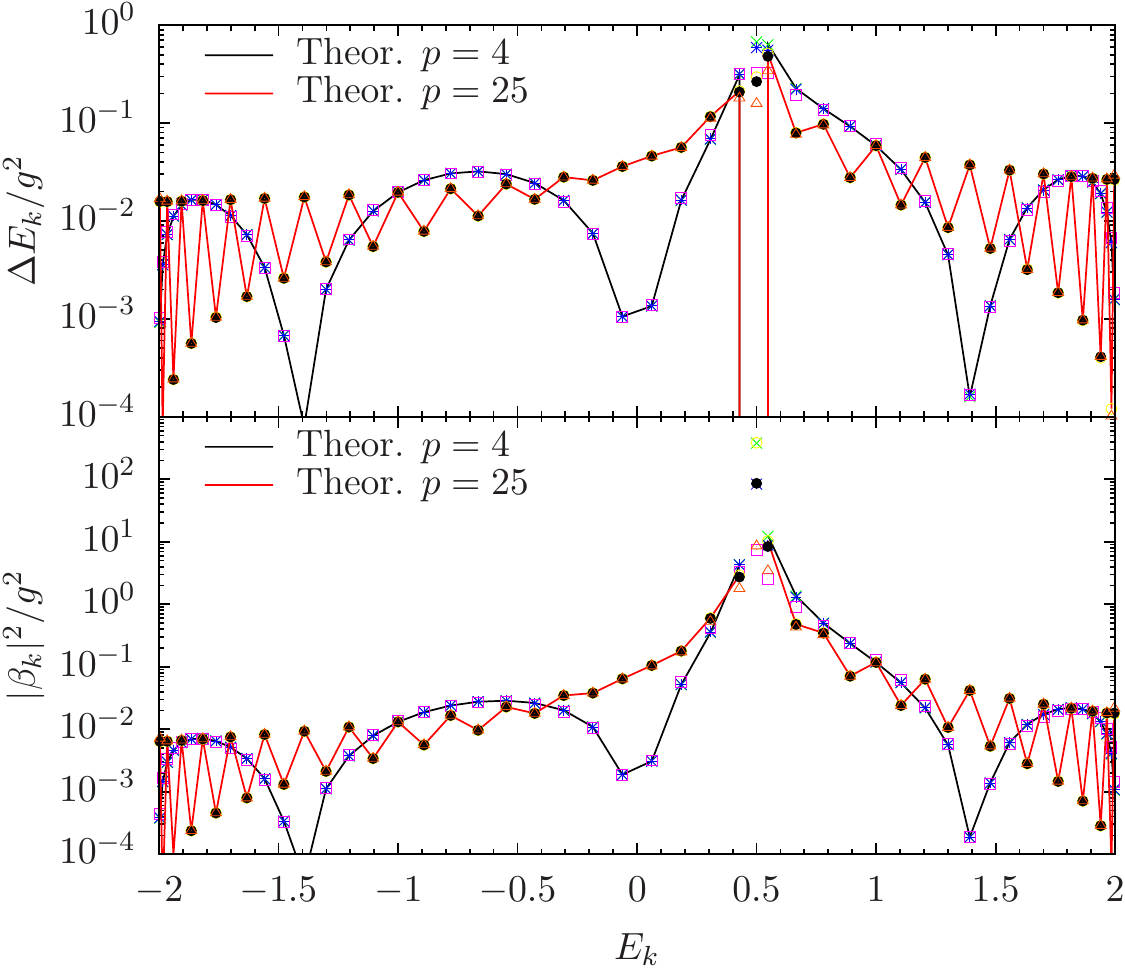}
\caption{Checking the validity of the two-level variational approach
  to single qubit detector physics. A hopping model with $L=50$ sites
  and open boundaries is attached to a thermometer at site $p=4$ (blue
  line) and $p=25$ (red line). Top: theoretical estimate for the shift
  in the mode energy $\Delta E_k=\tilde E_k- E_k$ due to the presence
  of the thermometer, as a function of $E_k$ and divided by $g^2$,
  Eq. \eqref{eq:deformed_energy}. The points are the exact values for
  $g=0.05$, $0.1$ and $0.25$, which are seen to collapse very
  accurately away from the region $E\sim \mu$. Bottom: theoretical
  estimate for the occupation of the thermometer site $|\beta_k|^2$ as
  a function of the unperturbed energy $E_k$, divided also by $g^2$,
  Eq. \eqref{eq:deformed_occupation}. The points are again the exact
  values for $g=0.05$, $0.1$ and $0.25$.}
\label{fig:sprout}
\end{figure}

The astonishing validity of this approximation can be seen in
Fig. \ref{fig:sprout}, where we compare the exact and the two-level
variational results with the exact calculation.

Returning to expression \eqref{eq:longtermaverage} we can state that
$|D_{0l}|^2=|\beta_l|^2$ and, approximately, $U_{kl}\approx
\delta_{lk}$, thus obtaining

\begin{equation}
n_0 \approx \sum_{k\in K} {g^2|\psi^k_p|^2\over (E_k-\mu)^2}.
\label{eq:longtermsprout}
\end{equation}

Notice that the local occupation (and, therefore, the local
temperature) is related to the form of the orbitals and the {\em
  energy content} at the site to which the detector is attached. The
long-term average occupation always depends quadratically with the
coupling constant, $n_0 \sim g^2$, for low enough $g$. Let us remark
again that in order to define a proper local temperature one should
always take the limit $g\to 0$. As we mentioned before, strictly
speaking one should distinguish between the {\em observed} local
temperature, at finite $g$, and the actual limit value, which is zero
everywhere for a ground state.

%%%%%%%%%%%%%%%%%%%%%%%%%%%%%%%%%%%%%%%%%%%%%%%%%%%%%%%%%%%%%%%%%%%%

\section{Conclusions and Further Work}
\label{sec:conclusions}

In this work we have presented an operational definition of the local
temperature of a quantum system, via the interaction with a single
qubit Unruh-DeWitt detector characterized as a two-level system with a
(large enough) energy gap $\mu$ and a (small enough) coupling constant
$g$. The main observable is the long-term average occupation of the
detector, which is shown to have a mild dependence on $\mu$ if it is
sufficiently above the Fermi energy \cite{finalremark}.

We have studied the behavior of the detector occupation and the
associated local temperature on the ground state of free fermionic
systems in 1D with inhomogeneous hopping parameters $t(x)$, which can
be understood as the time-lapse, $t(x) \sim |g_{00}(x)|^{1/2}$, of a
background static geometry (or local speed of light). Since we operate
at zero temperature, the energy to excite the detector must come only
from the coupling between the thermometer and the system, measured by
the coupling constant $g$. Indeed, the local thermometer occupation is
always proportional to $g^2$ and tends to zero as $g\to 0$. Thus,
properly speaking, the measured temperature is always zero. Yet, for
small but finite values of $g$ we find an approximate inverse
proportionality between the long-term average occupation of the
detector and the time-lapse, $n_0^{-1} \sim t(x)/g^2$. Thus, for
finite $g$, the {\em observed} local temperatures are larger where the
local speed of light is smaller, bearing similarity to the
Tolman-Ehrenfest theorem from thermodynamics on curved space-time,
which states that for a system in thermal equilibrium on a static
metric, $T(x) \cdot t(x)$ is a constant. Yet, in opposition to it, we
find that, for finite $g$, $T_g(x)\cdot\log t(x)\sim\text{const}$. Our
result does not contradict the Tolman-Ehrenfest theorem because the
limit of $T_g(x)$ for vanishing coupling $g$ is zero.

Our main claim, $n_0(x)^{-1} \sim t(x)/g^2$, seems to remain
approximately valid for a wide variety of inhomogeneities: linear
(Rindler), exponential (rainbow) or sinuosidal hoppings. Nonetheless,
a theoretical explanation and a discussion of its validity are left
for further work. It is relevant to ask whether it remains valid in
higher dimensions, in different topologies, or in the presence of
interactions.

The most relevant question is how our technique will work in the case
of equilibrium states at a finite temperature or non-equilibrium
systems. In that case, there are some relevant approaches in the
literature to define local and effective temperatures. A well tested
procedure is to attach a local thermal reservoir at temperature $T_0$
locally to the system, and find the value of $T_0$ for which the heat
flow between the bath and the system vanishes
\cite{Engquist.81,Dubi.09,Caso.10,Caso.11,Eich.16,Shastry.15}. This
approach is operational, like ours, and will also yield zero
temperature for the ground state. The main advantage of our approach
is that it explores the possibility of {\em measuring the temperature
  without a thermal bath}, and thus it is better suited for pure
quantum environments, such as ultracold atomic gases. Another tested
approach is based on the fluctuation-dissipation theorem
\cite{Cugliandolo.11}. The temperature is defined from the relation
between the response to an impulse perturbation and the correlation
function. Thus, as opposed to the previous case, it is not an
operational definition, and it presents several technical issues in
the quantum regime \cite{Foini.11}. Nonetheless, one of the most
relevant insights from the technique is that the temperature can be
frequency-dependent, a feature that can be obtained from our full
frequency occupation $\hat n_0(\omega)$, Eq. \eqref{eq:n0_fourier}. A
good extension of the definition of temperature should respect the
principles of thermodynamics. Both approaches mentioned above are
known to respect the second principle, but for our technique this is
still to be proved.

Although our procedure is inspired by the Unruh-De Witt detector, it
is also important to stress the difference between the local
temperature measured and the Unruh temperature. In order to observe
the Unruh effect, an observer will move with constant acceleration
through the Minkowski vacuum. From her point of view, this motion will
translate into a change of her metric, which will become
Rindler. Thus, as opposed to our case, she will observe the Minkowski
vacuum through the lens of a Rindler Hamiltonian, as shown in
\cite{Laguna.16}.

As a last remark, we would like to stress that our proposal for the
definition of the local temperature is operational, and therefore it
can lead to experimental observation. An interesting setting would be
using ultracold atoms on an optical lattice.

%%%%%%%%%%%%%%%%%%%%%%%%%%%%%%%%%%%%%%%%%%%%%%%%%%%%%%%%%

\begin{acknowledgments}
We would like to acknowledge very useful discussions with G. Sierra,
A. Celi, M. Lewenstein, O. Castro-Alvaredo, B. Doyon and
S.N. Santalla. This work was funded by grants FIS-2012-33642 and
FIS-2012-38866-C05-1, from the Spanish government, and by the Campus
of Excellence UAM+CSIC.
\end{acknowledgments}

%%%%%%%%%%%%%%%%%%%%%%%%%%%%%%%%%%%%%%%%%%%%%%%%%%%%%%%%%%%%%%%%%%


\begin{thebibliography}{10}
\bibitem{Footnote.1} There are some important caveats when attempting
  to apply the Tolman-Ehrenfest theorem to the Unruh effect, see
  \cite{Verch}.
\bibitem{finalremark} We would like to stress that, despite of our use
  of Unruh-De Witt detectors, our measurement is not of an Unruh
  temperature.
\end{thebibliography}
\end{document}